\documentclass[12pt,a4paper]{article}
\usepackage{amsmath}
\usepackage{amssymb}
\usepackage{graphicx}
\begin{document}
 \begin{titlepage}
  \begin{flushright}
IFUP--TH/2011-16-r\\
  \end{flushright}
~
\vskip .8truecm
 \begin{center}
\Large\bf
Late-time expansion in the semiclassical theory of the Hawking radiation
\end{center}
\vskip 1.2truecm
\begin{center}
{Pietro Menotti} \\
{\small\it Dipartimento di Fisica, Universit{\`a} di Pisa and}\\
{\small\it INFN, Sezione di Pisa, Largo B. Pontecorvo 3, I-56127}\\
{\small\it e-mail: menotti@df.unipi.it}\\
 \end{center}
\vskip 1.2truecm
\centerline{July 2011}

\vskip 1.2truecm

\begin{abstract}
We give a detailed treatment of the back-reaction effects on the Hawking
spectrum in the late-time expansion within the semiclassical approach to the
Hawking radiation. We find that the boundary value problem defining the 
action of the
modes which are regular at the horizon admits in general the presence of
caustics. We show that for radii less that a certain critical value $r_c$ no
caustic occurs for all values of the wave number and time 
and we give a rigorous lower bound
on such a critical value. 
We solve the exact system of non linear equations defining the motion, by an
iterative procedure rigorously convergent at late times. 
The first two terms of such an expansion give the $O(\omega/M)$ correction 
to the Hawking spectrum.
\end{abstract}

\end{titlepage}

\eject

\section{Introduction}

The semiclassical treatment of the Hawking radiation was introduced by Kraus
and Wilczek in \cite{KWI,KWII} after which several developments followed. The
main interest of the treatment is to provide a method to compute the
back-reaction effect of the radiation on the black hole, or in different words
a method which keeps into account the conservation of energy, an effect which
is completely ignored in the the external field treatment of the phenomenon
\cite{hawking,unruh,FH}. 
The main idea is to replace the free field modes of the
radiation by the semiclassical wave function of a shell of matter or
radiation which consistently propagates in the gravitational field generated
by the back hole and by the shell itself. The shell dynamics was studied in
detail in many papers (see \cite{FMP,KWI,KWII,FLW,FM,menottiI,menottiII} 
where also a more complete list of references is found). 
In the original semiclassical treatment 
\cite{KWI,KWII,KVK} the spectrum of the
Hawking radiation is extracted through the standard Fourier analysis of the
regular modes. Later such a treatment was related to the tunneling picture;
such an approach gave also rise to several proposals and to controversy
\cite{PW,padman,chowdhury,zerbini,akhmedov,akhmedova,zerbini1,zerbini2,
zerbini3,pizzi,mann} and \cite{vanzo} for a vast list of references.

We think that  the mode analysis is still the clearest 
and safest way to extract the results in the semiclassical approach. 

The present paper is devoted to a detailed analysis of the
construction of the semiclassical modes and their time Fourier transform. The
action related to the modes which are regular on the horizon is defined
through mixed boundary condition, i.e. a condition on the value of the
conjugate momentum at $t=0$ and a condition at time $t$ on the
coordinate $r$. While it is easy to prove that the variational problem in
which coordinates are given both at time $0$ and $t$ does not present
caustics, i.e. at most one motion satisfies the variational problem, we prove
that for the above mentioned mixed boundary condition problem in
general caustics arise i.e. in general more that one trajectory in phase space
satisfies the mixed boundary conditions.

Qualitatively this phenomenon is due to the fact that the time to reach the
final value $r_1$ of the radius is an increasing function the mass of the
black hole given an initial value $r_0$, but such initial value $r_0$ through
the 
condition on the initial momentum is also an increasing function of the mass
of the black hole thus giving rise to two counteracting effects.

On the other hand we prove that if the end point $r_1$, at which the time
Fourier analysis will be performed, is less that a critical value, 
caustics to not occur and we give also a rigorous lower bound on 
that critical value.

In the original paper \cite{KWI} it was argued that the semiclassical
approximation is expected to be valid, for not too large values of $r_1$.  
If we
stay below the critical value $r_c$ we are in the favorable situation of
absence of caustics where the semiclassical wave function is well defined. 
It is well know on the other hand that the time Fourier analysis 
gives result independent of $r_1$ \cite{KWI}.

We come now to the computation of the action as a function of $t$. Such a
problem corresponds to the solution of a system of two highly non linear
equations where the two unknown are the value $H$ of the hamiltonian, which 
even if a constant of motion depends on the time $t$ of the boundary 
condition, and the shell position at time $t=0$, $r_0$ which also depends 
on the mixed boundary conditions. 
In \cite{KWI} and \cite{menottiI} a discussion was given
of a simplified system of equations obtained by keeping only the most
singular terms in the exact equations; even if such a truncation does not 
remove
the high non linearity of the system it simplifies it a lot. Instead here we
treat the full exact system of equations. By introducing an implicit time
variable we show that for large times such a system of equations is
equivalent to an other non linear equation which can be solved by a convergent
iterative procedure. We show that the first two terms of the convergent
iterative procedure are sufficient to provide the leading spectrum of the
radiation and its back-reaction correction terms of order $\omega/M$.
The treatment is detailed enough to show directly how at late times, higher
and higher momenta of the regular modes on the horizon contribute. From the
general viewpoint confining ourselves to large times is not a limitation. The
reason, as well known, is that the Hawking radiation is a late-time
phenomenon. This feature was particularly enlightened in the treatment of
\cite{FH} where the Hawking radiation is extracted by the Fourier analysis of
the modes on a bounded space-time region fixed in space but translated at
asymptotically large times. In fact the persistence in time is one of the key
features of the black hole radiation.

The paper is organized as follows: sections \ref{choicegauge} 
to \ref{equationmotion} are devoted to the gauge
choice, to the description of the reduced action for the shell dynamics and
the ensuing equations of motion. In section \ref{caustics} 
we prove the existence of
caustics and give a rigorous bound on the value of the critical radius $r_c$ 
below which no
caustic develops. Section \ref{latetimeexpansion} 
is devoted to the
non linear system of equations related to the regular modes and to its
solution through 
a convergent iterative procedure. Section \ref{saddlepoint} 
deals with the saddle point
calculation of the Bogoliubov coefficients and discusses the region of
validity for such a procedure. In section \ref{conclusions} we give some 
concluding remarks.
In the Appendix we collect the most important formulae relative to the shell
dynamics. 
As it is usual in this field, we work with $c=1$ and $G=1$; 
it means that time, momenta, mass and energy are all measured 
in units of length.

\section{Choice of gauge and the conjugate momentum }\label{choicegauge}

In the general expression of the metric
\begin{equation}\label{metric}
ds^2 =-N^2 dt^2+L^2(dr+N^r dt)^2 +R^2 d\Omega^2
\end{equation}
all quantities $N,L,N^r,R$ are supposed functions only of the radial variable 
$r$ and $t$ thus realizing spherical symmetry. 

The semiclassical approach is best developed in the Painlev\'e-Gullstrand
metric characterized by setting $L=1$ in (\ref{metric}).
Such a metric has the advantage of being non singular at the horizon. 
After fixing $L=1$ one has still a gauge choice on $R$. In
presence of a shell of matter one cannot choose $R=r$. One has
several choices; for a discussion see \cite{FM,menottiI,menottiII}. In the
present paper we will use the ``outer gauge'' which is defined by $R=r$ for
$r\geq \hat r$ where $\hat r$ denotes the shell position. At $r=\hat r$, 
$R$ is continuous as all the other functions appearing in (\ref{metric}), 
but its
derivative is discontinuous. For the reader's convenience we report in the
Appendix the main results on the shell dynamics which are necessary in the 
following developments.

The first step is to go over from the standard Hilbert-Einstein action added
to the action of the matter shell, to the action expressed in Hamiltonian
form.

As usual in gravity it is better to work on a bounded region of
space-time. The radial coordinate will range from $r_i$ to $r_e$ while time
ranges from $t_i$ to $t_f$.

After solving the constraints one can rigorously express the action in reduced
form i.e. a form in which only the coordinate $\hat r$ of the shell and a 
conjugate momentum appears, in addition to the boundary terms. 
As always these are essential 
in gravity, where the boundary terms play the role of the Hamiltonian. The
reduced action in the outer gauge is given by \cite{KWI,FM,menottiI} 
(see also the Appendix)
\begin{equation}\label{reducedaction}
S = \int_{t_i}^{t_f} \left(p_{c}~ \dot{\hat r} -\dot M(t)\int_{r_i}^{\hat
r(t)}\frac{\partial F}{\partial M} dr - H N(r_e) + M N(r_i)\right)dt 
\end{equation}
where $F$ is the generating function
\begin{eqnarray}\label{Ffunction}
& &F=R \sqrt{(R')^2 -1+ \frac{2{\cal M}}{R}} \\
&+&
RR'\left(\log(R'- \sqrt{(R')^2 -1+ 
\frac{2{\cal M}}{R}})-\log \big(1-\sqrt\frac{2 \cal M}{R}\big)-
\sqrt{\frac{2{\cal M}}{R}}\right).\nonumber
\end{eqnarray}
As a consequence of the constraints the quantity ${\cal M}$ which appears in
(\ref{Ffunction}) is constant in $r$ except at the position of the shell where
it is subject to a discontinuity. 
$M$ and $H$ are the the value of the quantity ${\cal M}$ below and above the
shell position and thus at $r_i$ and $r_e$.
One can consider either $M$ or $H$ as a given datum
of the problem. In the outer gauge for which the action has the form
(\ref{reducedaction}) it is simpler to consider $M$ as a datum of the problem
which to be consistent with the gravitational equations has to be constant in
time \cite{FM}; so the term proportional to $\dot M$ disappears and we reach
\begin{equation}\label{reducedaction2}
S = \int_{t_i}^{t_f} \big(p_{c}~ \dot{\hat r} - H N(r_e) + M N(r_i)\big)dt .
\end{equation}
As in the variation, the components of the metric have to be kept constant at
the boundaries, action (\ref{reducedaction2}) with the normalization $N(r_e)=1$ 
is equivalent to 
\begin{equation}\label{reducedaction3}
S = \int_{t_i}^{t_f} \big(p_{c}~ \dot{\hat r} - H)dt .
\end{equation}
Even though the kinetic momentum of the shell $\hat p$ (see Appendix) is a 
gauge dependent quantity,
the conjugate momentum $p_c$ appearing in the reduced action is a gauge
invariant quantity \cite{menottiI}.  
It can be computed both for a massive or massless shell
\cite{KWI,FLW,FM,menottiI}; as we shall in this paper be interested in the 
massless case we report below its expression only for the massless case
\begin{equation}\label{pc}
p_c =\sqrt{2M \hat r}-\sqrt{2H \hat r}-\hat r \log\frac{\sqrt{\hat
r}-\sqrt{2H}}{\sqrt{\hat r}-\sqrt{2M}}.
\end{equation}    
One has to keep in mind that $p_c$ is not the kinetic momentum of the shell
but the conjugate momentum with respect to $\hat r$ of the whole system. 

\section{The action for the modes regular at the horizon}\label{regularaction}

As in the following of the paper only $\hat r$ would appear, we will for
notational simplicity denote $\hat r$ (the shell position) simply by $r$
without any possibility of confusion.

\bigskip

At the semiclassical level the modes which are invariant under the Killing
vector $\frac{\partial}{\partial t}$ are simply given by

\begin{equation}
e^{iS/l_P^2}
\end{equation}    
with $l_P^2= G\hbar$ the square of the Planck length and
\begin{equation}\label{killingaction}
S = \int^{r_1} p_c dr - Ht +{\rm const}.
\end{equation}    
As is well know such modes have the feature of being singular at the horizon;
this is immediately seen from the expression of $p_c$ eq.(\ref{pc}) which
diverges at $r= 2H$. The vacuum given by $a_\omega |0\rangle=0$, being
$a_\omega$ the destruction operator relative to the described modes 
gives rise to a singular description at the horizon, while a free falling
observer should not experiment any singularity \cite{unruh}.

Instead the true vacuum should be described in term of modes which are regular
at the horizon \cite{unruh,KWI}. Thus the main object which intervenes in the
semiclassical treatment is the semiclassical expression of the modes 
regular at the horizon i.e. the action of the system which describes an
outgoing shell of matter and has the following boundary conditions \cite{KWI}: 
i) at time $0$ the conjugate momentum is a given value $k$; 
ii) at time $t$ the shell position $r$ is a given value $r_1$. The expression
for such an action was already given by Kraus and Wilczek in \cite{KWI}.
With the two conditions $p_c(0) =k$ and $r(t) = r_1$ the  action is
\begin{eqnarray}\label{KWaction}
S(r_1,t,k) &=& k r_0(r_1,t,k) +\int_0^{t}\big(p_c \dot{r} - 
H(r(t'),p_c(t'))\big) dt'=\nonumber\\
&=& k r_0(r_1,t,k) +\int_0^{t}p_c \dot{r} dt' - 
H[r_1,t,k] t.
\end{eqnarray}
The last equality is due to the fact that $H$ along the motion
is a constant despite $H$ depends on the boundary conditions as
explicitely written. $r_0$ denotes the value of $r$ at time $0$; also such a
quantity depends on the imposed boundary conditions. 
Taking into account that $r$ and $p_c$ depend both on the final time $t$ and
on the running time $t'$, and denoting with a dot the derivative with respect
to $t'$ one has
\begin{equation}
\frac{\partial S}{\partial r_1} = k\frac{\partial r_0}{\partial r_1}
+\int_0^{t}\big(p_c\frac{\partial \dot r}
{\partial r_1}+\dot p_c\frac{\partial r}{\partial r_1}\big)dt'=p_c.
\end{equation}
Similarly
\begin{equation}\label{dSodt}
\frac{\partial S}{\partial t} = k\frac{\partial r_0}{\partial t}
+(p_c \dot r-H)|_t+ \int_0^t\big(p_c\frac{\partial \dot r}{\partial t}
+\dot p_c\frac{\partial r}
{\partial t}\big)dt'=- H.
\end{equation}
The action (\ref{KWaction}) has to be computed on the solution of the
equation of motion,satisfying the described boundary conditions.

\section{The equations of motion}\label{equationmotion}

In the outer gauge \cite{FM,menottiI}, that we adopt here, the equation 
of motion for $r$ has the form
\begin{equation}\label{dotr}
\frac{dr}{dt}=1-\sqrt\frac{2H}{r}
\end{equation}
while $dp_c/dt$ can be obtained substituting $\dot r(t)$ in eq.(\ref{pc}).
Eq.(\ref{dotr}) can be integrated in the form
\begin{equation}\label{tequation}
t=
4H\log\frac{\sqrt{r_1}-\sqrt{2H}}{\sqrt{r_0}-\sqrt{2H}}+r_1-r_0+2\sqrt{2Hr_1} 
-2\sqrt{2Hr_0}.
\end{equation}
The boundary condition at $t=0$ gives
\begin{equation}\label{kequation}    
0<k =\sqrt{2M r_0}-\sqrt{2H r_0}- r_0 \log\frac{\sqrt{r_0}-\sqrt{2H}}
{\sqrt{r_0}-\sqrt{2M}}=\int_M^H\frac{dH'}{1-\sqrt\frac{2H'}{r_0}}
\end{equation}    
where $2M<2H<r_0<r_1$; eq.(\ref{kequation})
together with eq.(\ref{tequation}) should determine completely the
motion. However as we mentioned in the introduction we shall find that for
sufficiently large $r_1$ caustics arise, i.e. there exist values of $t$, $r_1$
and $k$ for which the boundary conditions are satisfied by more than one
motion. 

At $t=0$ we have $S(r_1,0,k) = kr_1$ which is regular at the horizon and in
virtue of the equations of motion $S$ remains regular in the time development.

\section{The occurrence of caustics}\label{caustics}

It is very easy to show that the standard variational problem in which $r$ is
fixed to $r_0$ at time $0$ and to $r_1$ at time $t$ presents no caustics. In
fact from
\begin{equation}\label{timeequation}    
t=\int_{r_0}^{r_1}\frac{dr}{1-\sqrt\frac{2H}{r}}
\end{equation}      
we see that $t$ is an increasing function of $H$. Thus there is at most one
$H$ which satisfies the boundary conditions. But the value of $r$
at $t$ and $H$ determine completely the motion for an outgoing shell. On
the other hand the problem (\ref{KWaction}) which has mixed boundary 
conditions is more complicated. 
First we note that being, from eq.(\ref{kequation}) 
\begin{equation}\label{dkdH}
\bigg(\frac{\partial k}{\partial
H}\bigg)_{r_0}=\frac{1}{1-\sqrt\frac{2H}{r_0}}>0     
\end{equation}      
we have that at fixed $k$, $H$ is a single valued function of $r_0$ and
viceversa $r_0$ a single valued function of $H$.
Moreover we have
\begin{equation}
\bigg(\frac{\partial k}{\partial r_0}\bigg)_H =- 
\int^H_M\frac{dH'}{2r_0(1-\sqrt{\frac{2H'}{r_0}})^2}
\sqrt{\frac{2H'}{r_0}}=-\frac{1}{2}
\int^{\sqrt{a}}_{\sqrt\frac{2M}{r_0}}\frac{y^2dy}
{(1-y)^2}<0
\end{equation}
with
\begin{equation}
a=\frac{2H}{r_0}.
\end{equation}      
Combined with eq.(\ref{dkdH}) it gives
\begin{equation}
\bigg(\frac{\partial H}{\partial
r_0}\bigg)_k=\frac{1}{2}(1-\sqrt{a})
\int^{\sqrt a}_{\sqrt\frac{2M}{r_0}}\frac{y^2dy}{(1-y)^2}>0.
\end{equation}      
To investigate the occurrence of caustics we shall compute the derivative of
$t$ with respect to  $r_0$ under the constraint of constant $k$.
First we note that from eq.(\ref{timeequation})
\begin{equation}
\bigg(\frac{\partial t}{\partial H}\bigg)_{r_0}=\int_{r_0}^{r_1}\frac{dr'}
{(1-\sqrt\frac{2H}{r'})^2}
\sqrt\frac{2}{r'}\frac{1}{2}\frac{1}{\sqrt{H}}=
2\int_{\sqrt\frac{2H}{r_1}}^{\sqrt\frac{2H}{r_0}}\frac{dz}{z^2(1-z)^2}>0
\end{equation}
and
\begin{equation}    
\bigg(\frac{\partial t}{\partial r_0}\bigg)_H=-\frac{1}{1-\sqrt a}.
\end{equation}      
Thus
\begin{equation}\label{dtdr0}    
\bigg(\frac{\partial t}{\partial r_0}\bigg)_{k}=
\bigg(\frac{\partial t}{\partial
r_0}\bigg)_H+\bigg(\frac{\partial t}{\partial H}\bigg)_{r_0}
\bigg(\frac{\partial H}{\partial r_0}\bigg)_k= -\frac{1}{1-\sqrt a}[
1 - I_1 ~I_2]
\end{equation}      
with
\begin{equation}    
I_1= (1-\sqrt a)\int^{\sqrt{a}}_{\sqrt\frac{2M}{r_0}}
\frac{y^2 dy}{(1-y)^2}~,~~~~~~~~
I_2= (1-\sqrt a)\int_{\sqrt\frac{2H}{r_1}}^{\sqrt{a}}\frac{dz}{z^2(1-z)^2}.
\end{equation}
It is easily seen that
\begin{equation}    
0<I_1\leq 1,~~~~0<I_2.
\end{equation}
The value of eq.(\ref{dtdr0}) for $r_0=r_1$, due to the vanishing of $I_2$, 
is the finite negative value
\begin{equation}\label{dtsdr00}    
\bigg(\frac{\partial t}{\partial r_0}\bigg)_k(r_1)=
-\frac{1}{1-\sqrt\frac{2H_1}{r_1}}<0
\end{equation}
with $H_1$ solution of eq.(\ref{kequation}) in which $r_0$ has 
been substituted with
$r_1$. On the other hand given a value of $k$ and of $r_0$ 
(which through eq.(\ref{kequation}) 
gives a value of  $H$ with $2M<2H<r_0$), 
there will always be $r_1$ large enough as to make 
the product
$ I_1 ~I_2$ larger that $1$; this because $I_2$ diverges when the lower
integration limit goes to zero. Thus at that point eq.(\ref{dtdr0}) 
becomes positive while (\ref{dtsdr00}) still has to hold.
Summarizing we found that for a given $k$, for large enough $r_1$ 
the derivative (\ref{dtdr0}), when $r_0$ moves from $r_1$ to $2M$
changes sign,
thus vanishing at at least one intermediate point. This implies the
occurrence of caustics \cite{arnold}. 
In fact the vanishing of the derivative (\ref{dtdr0}) at the
value $r_0^*$ implies that there will be points $r_0'$  and $r_0''$
on the right and on the left of $r^*_0$ which give rise to the same value of
$t$. Thus we shall have pairs of distinct motions with the same $k$ which reach
$r_1$ at the same time $t$. (One can also give numerical examples of such
pairs of motions). In constructing caustics we took $r_1$ large
enough. We will show now that for $r_1<r_c$ where $r_c$ is a critical value,
no caustic arises, for any $k$.

Below we give a simple procedure to give a rigorous lower bound on $r_c$.
It is very simple to show that  for $\sqrt a=1$ both $I_1$
and $I_2$ are equal to $1$. Setting
\begin{equation}    
I_1 = 1+\Delta_1,~~~~~~~~I_2 = 1+\Delta_2
\end{equation}
we will prove that
\begin{equation}\label{suminequality} 
\Delta_1 + \Delta_2 <0
\end{equation}
for $r_1$ less than a value $r_b$ independently of the value of $k$. Then being
$I_1>0$ and $I_2>0$ we have $I_1~I_2<1$ and thus $(\frac{\partial t}{\partial
r_0})_k$ always negative.
 
Thus for $r_1<r_b$ there will be no caustic i.e. $r_b$ will
constitute a lower bound on $r_c$.  With regard to the proof of
(\ref{suminequality}) explicit computation of the integrals gives
\begin{eqnarray}\label{inequality} &&\frac{\Delta_1 + \Delta_2}{1-\sqrt a} =
\sqrt a -\frac{1}{1-\sqrt\frac{2M}{r_0}}-\sqrt\frac{2M}{r_0}
-2\log\bigg(1-\sqrt\frac{2M}{r_0}\bigg)\nonumber\\ 
&-&\frac{1}{\sqrt
a}-2\log\sqrt\frac{r_0}{r_1}- \frac{1}{1-\sqrt\frac{r_0}{r_1}\sqrt a}+
\frac{1}{\sqrt\frac{r_0}{r_1}\sqrt a} +2 \log\bigg(1-\sqrt\frac{r_0}{r_1}\sqrt
a\bigg)\leq\nonumber\\ 
&&-1-2\log\sqrt\frac{r_0}{r_1}-
\frac{1}{1-\sqrt\frac{r_0}{r_1}\sqrt a}+ \frac{1}{\sqrt\frac{r_0}{r_1}\sqrt a}
+2 \log\bigg(1-\sqrt\frac{r_0}{r_1}\sqrt a\bigg)\leq\nonumber\\
&&-1-2\log\sqrt\frac{2M}{r_1}-\frac{1}{1-\sqrt\frac{2M}{r_1}}
+\frac{1}{\sqrt\frac{2M}{r_1}}+2\log\bigg(1-\sqrt\frac{2M}{r_1}\bigg)
\end{eqnarray} 
where in writing the two inequalities we used repeatedly $2M\leq
2H\leq r_0$.  The last term in eq.(\ref{inequality}) is a decreasing function
of $2M/r_1$ and it is less that zero for $2M/r_1=2/10$. Thus we do not have
caustics for $r_1<10 M$ and as a consequence 
we have rigorously $r_c>10 M$ for the critical
value $r_c$.  A numerical search of eq.(\ref{dtdr0}) gives the wider bound
$r_c> 24 ~M$.  In \cite{KWI,menottiI} the approximate system of equation
obtained by retaining in eqs.(\ref{tequation},\ref{kequation}) only the
singular terms i.e. only the logarithms was considered. Also for this
approximate system of equations, caustics occur for $r_1$ sufficiently
large. The occurrence of caustics for $r_1>r_c$ hints at a failure of
the 
semiclassical approximation when we move too far from the horizon as the modes
would show a discontinuity in $r_1$ at the point where more than one
trajectory in phase space start contributing. On the
other hand we will show in section \ref{latetimeexpansion}, 
that for any given pair
$(r_1,k)$ even for $r_1>r_c$ for $t$ sufficiently large no caustic occurs.

In \cite{KWI} it was proposed to perform the time Fourier analysis at a point
$r_1$ not too far from the horizon, the reason being that there one should
expect the semiclassical approximation to be reliable.  We showed above that
for $r_1<r_c$ there are no ambiguities in the definition of the action and in
addition it is well known the time Fourier transform gives results 
independent of
$r_1$; thus we shall work with $r_1<r_c$.

\bigskip

\section{The late-time expansion}\label{latetimeexpansion}

In this section we shall give the solution of the equations for $H(t)$ and
$r_0(t)$ in the
form of a series convergent for large times. Large times are also the
relevant times for the treatment of the Hawking radiation.
We recall that $2M<2H<r_0<r_1$. Then from eq.(\ref{tequation}) 
for $r_1$ fixed, $t\rightarrow+\infty$ implies
\begin{equation}    
\sqrt{r_0}-\sqrt{2H}\rightarrow 0.
\end{equation}
Looking now at eq.(\ref{kequation}) we must have in the same limit
\begin{equation}    
\sqrt{r_0}-\sqrt{2M}\rightarrow 0
\end{equation}
and as $2M<2H<r_0$ we have also $H\rightarrow M$.
We introduce now the implicit time variable $T=\exp(-{\frac{t}{4H}})$ 
which due to the bounds on $H$, for $t\rightarrow +\infty$ tends to
$0$. Eq.(\ref{tequation}) becomes 
\begin{equation}\label{Tequation}
T\equiv e^{-\frac{t}{4H}}=\frac{\sqrt{r_0}-\sqrt{2H}}{\sqrt{r_1}-\sqrt{2H}}
\exp\bigg(-\frac{r_1-r_0}{4H}-\sqrt\frac{r_1}{2H}+\sqrt\frac{r_0}{2H}\bigg). 
\end{equation}
It will be useful for the following developments to use the notation
\begin{equation}\label{notation}    
h = \sqrt{2H} ,~~~~m = \sqrt{2M},~~~~v_0= \sqrt{r_0},~~~~ A = \sqrt{r_1}-m>0
\end{equation}
and set   
\begin{equation}
h-m= T c_H,~~~~v_0-m= T c_R
\end{equation}
with $c_H$ and $c_R$ functions of $T$ do be determined.
Eq.(\ref{Tequation}) becomes $F_1 = 1$ with
\begin{equation}\label{F1}
F_1= \frac{c_R-c_H}{A- T c_H} 
\exp\big[-\frac{(A-c_RT)(A+4m+(2c_H+c_R)T)}{2(m+c_H T)^2}\big]
\end{equation}
and
eq.(\ref{kequation}) becomes $F_2= k$ with
\begin{equation}\label{F2}
F_2 = -c_H T(m+c_R T)-(m+ T c_R)^2 \log\frac{c_R - c_H}{c_R}.
\end{equation}
We want to express $h$ as a function of the implicit variable $T$.

First we note that for $T=0$ the system of the two equations 
$F_1=1,~F_2=k$ has the unique solution
\begin{equation}    
c_H^0 = (e^\frac{k}{m^2}-1)\frac{A}{E},~~~~c_R^0 = e^\frac{k}{m^2}\frac{A}{E}
\end{equation}
where 
\begin{equation}    
E = \exp\big(-\frac{A(A+4m)}{2m^2}\big)
\end{equation}
and that $F_1$ and $F_2$ in a polydisk around $T=0$, $c_H=c_H^0$, $c_R=c_R^0$
are analytic functions of $T,c_H,c_R$. We have
\begin{equation}    
\frac{\partial F_2}{\partial c_R}\bigg|_{0, c_H^0,c_R^0}= -m^2
\frac{1}{c_R^0-c_H^0}\frac{c^0_H}{c^0_R}\neq 0
\end{equation}
and thus according to the implicit function theorem \cite{GR}, in a
neighborhood of $c_R^0$, $c_R$ will be an analytic function of $T$ and $c_H$.
Substituting in $F_1$ we obtain the equation
\begin{equation}    
1 = F_1(T, c_H, c_R(T,c_H)).
\end{equation}
At $T=0$, $c_H=c_H^0$ we have
\begin{equation}
\frac{\partial F_1}{\partial c_H}\bigg|_{0,c_H^0}=     
\frac{\partial F_1}{\partial c_H}+\frac{\partial F_1}{\partial c_R}
\frac{\partial c_R}{\partial c_H} = 
\frac{E}{A}\big(-1+\frac{c_R^0}{c_H^0}\big)=\frac{E}{A}~\frac{1}
{e^\frac{k}{m^2}-1}\neq 0
\end{equation}
and thus $c_H$ will be an analytic function $f(T)$ of $T$ in a neighborhood 
of $T=0$.
Recalling now the definition of $c_H$ we have
\begin{equation}\label{implicit}    
h = m+T f(T).
\end{equation}
Summing up we found that in a
neighborhood of $T=0$ from the equations $F_1=1,~F_2=k$ 
eq.(\ref{implicit}) follows. Eq.(\ref{implicit})
due to the definition of $T=\exp(-t/(2h^2))$ is still an implicit equation. 
We show now that eq.(\ref{implicit}) 
can be solved by a convergent iterative procedure.

Being $f(T)$ analytic with $f(0)=c^0_H>0$,
$g(t)=Tf(T)$ will be in a neighborhood of $T=0,T\geq 0$ 
a non negative function of $T$
of Lipschitz type, i.e. $0<|g(T_2)-g(T_1)| <c|T_2-T_1|$ for $T_1,T_2$ 
belonging to such a neighborhood. Moreover $g(0)=0$.
For $t$ such that
\begin{equation}\label{injection}
c e^{-\frac{t}{2 r_1}}<A    
\end{equation}
the r.h.s. of eq.(\ref{implicit}) maps the domain $m<h<A+m = \sqrt{r_1}$ into
itself. We start the 
iterative process with $h_0=m$. We must give a bound on $|h_{n+1}-h_{n}|$.
We have for $h_n-h_{n-1}>0$
\begin{equation}    
|h_{n+1}-h_{n}|\leq c e^{-\frac{t}{2h^2_n}}(1 -e^{-\frac{t}{2}
(\frac{1}{h^2_{n-1}}-\frac{1}{h^2_n})})\leq c e^{-\frac{t}{2h^2_n}}
~t~\frac {h_n-h_{n-1}}{h_n h^2_{n-1}}\leq \frac{c}{m} \frac{t}{m^2}
e^{-\frac{t}{2r_1}}
|h_n-h_{n-1}|.
\end{equation}
and for $h_n-h_{n-1}<0$ we reach the same result. Thus for $t$ satisfying
eq.(\ref{injection}) and 
\begin{equation}    
\frac{c}{m} \frac{t}{m^2}e^{-\frac{t}{2r_1}}<1
\end{equation}
we have a contraction mapping and according to Banach fixed point theorem
\cite{deimling} eq.(\ref{implicit}) has one and only one solution given by the
convergent sequence $h_n$. 

We give in Fig.1 a qualitative graph of the behavior in time of $2H(t)$ and
$r_0(t)$.
\begin{figure}[htb]
\begin{center}
\includegraphics{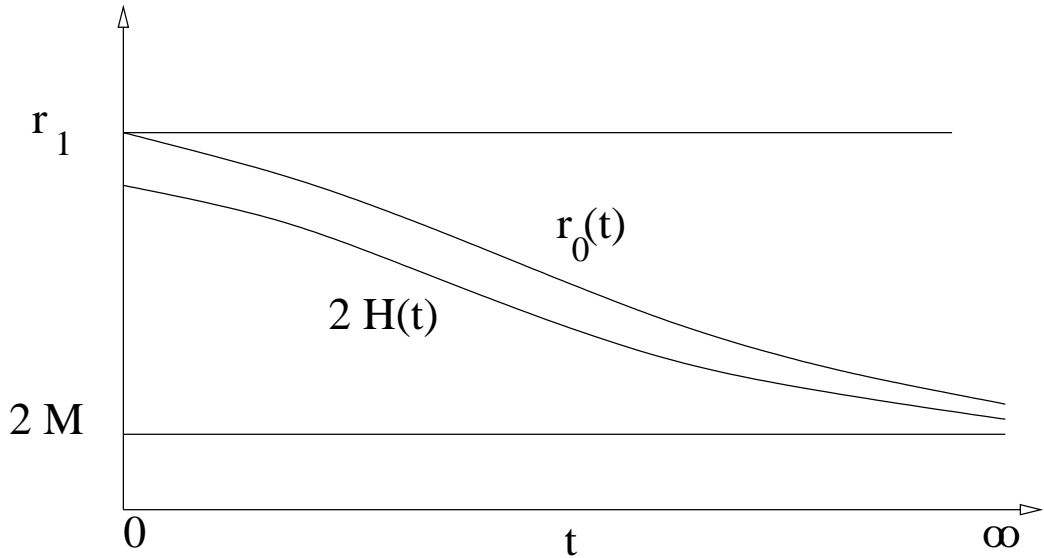}
\end{center}
\caption{Time development of $2H(t)$ and $r_0(t)$}\label{fig:1}
\end{figure}

We work out now explicitely the first two terms of such an iteration
procedure; they will be sufficient to give the $O(\omega/M)$ corrections to
the Hawking distribution. With $h_0=m$ and $\tau \equiv e^{-t/(4M)} = 
e^{-t/(2 m^2)}$ we have
\begin{equation}
h_1 = m +c_H^0 \tau,~~~~ h_2 =m+c_H^0\tau +\frac{t}{m^3} (c_H^0\tau)^2
+O(\tau^2)
\end{equation}
and thus for $H(t)$
\begin{equation}\label{H(t)}
H(t) = M +\sqrt{2M}c_H^0\tau +\frac{t}{2M}(c^0_H\tau)^2+O(\tau^2).    
\end{equation}
Due to eq.(\ref{dSodt}) the time dependence of the mode which is 
regular at the horizon, for fixed $r_1$  is
\begin{equation}
-\int^t H(t') dt' = {\rm const}~- Mt +4 M\sqrt{2M} \tau_1 + t
\tau_1^2 + O(\tau_1^2)    
\end{equation}
with $\tau_1 \equiv c_H^0 \tau$
i.e. for the semiclassical mode we have
\begin{equation}\label{actionexp}
e^{iS/l_P^2}
= e^{i [q(r_1) - Mt +4 M\sqrt{2M} \tau_1 + t \tau_1^2 + O(\tau_1^2)]/l_P^2}    
\end{equation}
where $l_P^2 = G\hbar$ is the square of the Planck length.
Thus $S$ at large times behaves as $-Mt$ independently of $k$. 
On the other hand the Fourier time
analysis of $e^{iS/l_P^2}$ contains frequencies which are above and below the
value 
$M$ and this is the well known fact that the mode of the system which is
regular at the horizon does not represent an eigenvalue of the energy as
measured by a stationary observer at space infinity.  The deviations from the
value $M$ represent the positive and negative frequency content of the
radiation mode. One has to keep in mind that the action which appears in
(\ref{actionexp}) 
refers to the whole system, which includes both the shell and the core.
If we want to analyze the modes of the radiation we have to subtract from the
exponent the background term $-Mt$.

\section{The saddle point approximation}\label{saddlepoint}

As well known and discussed in \cite{KWI,menottiI} the Bogoliubov
coefficients $\alpha_{\omega k}$ and $\beta_{\omega k}$ are given by 
\begin{equation}\label{alphabeta}    
\alpha_{\omega k}= c(r_1) \int dt ~e^{i (S + Mt +\omega t)/l_P^2},~~~~
\beta_{\omega k}= c(r_1) \int dt ~e^{i (S + Mt -\omega t)/l_P^2}.
\end{equation}
As discussed in section \ref{caustics} we will work with $r_1<r_c$.
The above integrals will be computed using the saddle point method where 
$l_P^2$ plays the role of asymptotic parameter
\cite{erdelyi}. From what we derived in the previous section, the exponent
appearing in the integrands, multiplied by $-i l_P^2$ apart from $q(r_1)$ 
which is constant in time and common to both coefficients, are respectively
\begin{equation}\label{reducedexponent}
2 m^3\tau_1 +t\tau_1^2+m^2(s+1)\tau_1^2 \pm \omega t~~~~{\rm with}~~
\tau_1 = c^0_H \tau
\end{equation}
where we used the notation of eq.(\ref{notation}), 
For the $\alpha_{\omega k}$ case (i.e. upper sign) the saddle point is given 
by the 
value of time $t$ which satisfies
\begin{equation}\label{alphasaddlepoint}
0= -H(t)+M+\omega=-m\tau_1-\frac{t}{m^2}\tau_1^2-s\tau_1^2+\omega
\end{equation}
which being $\omega>0$ has solution for real $t$ and thus at a real value of
the exponent in eq.(\ref{alphabeta}). On the contrary for the
$\beta_{\omega k}$ case (lower sign), the saddle point equation 
\begin{equation}\label{betasaddlepoint}
0= -H(t)+M -\omega=-m\tau_1-\frac{t}{m^2}\tau_1^2-s\tau_1^2-\omega
\end{equation}
has solution for complex $t$. At such a value of time the exponent
(\ref{reducedexponent}) (lower sign) equals 
\begin{equation}\label{betaexponent}
B = -2m^2\omega -t(\tau_1^2+\omega)-(s-1)m^2\tau_1^2.   
\end{equation}
The solution of eq.(\ref{betasaddlepoint}) to second order in $\omega$, 
which is the order we are interested in, is given by
\begin{equation}\label{tau1solution}
\tau_1 =
-\frac{\omega}{m}\bigg(1-\frac{2\omega}{m^2}\log(-\frac{\omega}{c^0_Hm})+
\frac{s\omega} {m^2}\bigg).    
\end{equation}
From eq.(\ref{betaexponent}) we see that to find the imaginary part of such
exponent to order $\omega^2$ we simply need the imaginary part of $t$ to 
first order in $\omega$. Using (\ref{tau1solution}) we have
\begin{equation}
{\rm Im}~ t =-2\pi m^2(1-\frac{2\omega}{m^2}).    
\end{equation}
Substituting into eq.(\ref{betaexponent}) we find
\begin{equation}\label{ImB}
{\rm Im}~B  = 2\pi m^2 \omega(1-\frac{\omega}{m^2}) =4\pi M
\omega(1-\frac{\omega}{2M})     
\end{equation}
which according to (\ref{alphabeta}) has to be divided by $l_P^2$.
Thus we have
\begin{equation}\label{beta2oalpha2}
\frac{|\beta_{\omega k}|^2}{|\alpha_{\omega k}|^2} = e^{-8\pi
\frac{M\omega}{l_P^2}(1-\frac{\omega}{2M})} 
\end{equation}
which is independent of $k$. We see from eq.(\ref{H(t)}) that for $t\rightarrow
+\infty$, $H(t)$ tends to $M$ and thus the time Fourier transform of the
exponential of the action (\ref{actionexp}) which refers to the whole system 
has a singularity at the frequency
$M$. Recalling 
that $H$ (outer mass) represents the energy of the whole system, we identify
the parameter $M$ with the mass of the black-hole before the decay.

Using the property of the Bogoliubov coefficients 
\begin{equation}
\sum_k(\alpha_{\omega k} \alpha^*_{\omega' k}-
\beta_{\omega k} \beta^*_{\omega' k}) = \delta_{\omega,\omega'}
\end{equation}
one reaches for the flux of the Hawking radiation
\cite{KVK} 
\begin{equation}
F(\omega) d\omega =\frac{d\omega}{2\pi}\frac{1}{e^{8\pi
\frac{M\omega}{l_P^2}(1-\frac{\omega}{2M})}-1}~.
\end{equation}
This completes the explicit derivation of the $\omega^2$ correction to the
Hawking formula from the time Fourier transform of the semiclassical modes.

\bigskip

An alternative way to derive (\ref{beta2oalpha2}) was given 
by Keski-Vakkuri and Kraus \cite{KVK} where it is proven that for the
$\beta_{\omega k}$ coefficient the imaginary part of the action at the saddle
point (\ref{betasaddlepoint}) is given by
\begin{equation}\label{KVKrelation}    
{\rm Im} \int^{r_1}_{r_0} p_c dr ={\rm Im} \int^{2M}_{2H} p_c dr=
\pi\frac{1}{2}((2M)^2 -(2H)^2) = 4 \pi M\omega(1-\frac{\omega}{2M})   
\end{equation}    
which is equivalent to eq.(\ref{ImB}). The importance of equation 
(\ref{KVKrelation}) is to show
directly how the ``tunneling'' is due only to the imaginary part of the
``space part'' of the action.

With regard to the validity of the expansion we see from the saddle point 
value (\ref{alphasaddlepoint},\ref{betasaddlepoint}) 
\begin{equation}\label{explicit}
\frac{A(e^{\frac{k}{2M}}-1)}{\sqrt{2M}E}e^{-\frac{t}{4M}}\approx 
\frac{\omega}{2M}
\end{equation}
that the series if effectively an expansion in $\omega/M$ and thus expected to
hold for $\omega/M <<1$. From eq.(\ref{explicit}) we see that for a given
$\omega$, large values of the wave number $k$ contribute at times $t$ which 
grow like $2k$. The typical $\omega$ for the radiation emitted by a black hole
of mass $M$ is according to eq.(\ref{beta2oalpha2}) (Wien's law)
\begin{equation}\label{typicalfreq}
\omega \approx \frac{l_P^2}{8\pi M}
\end{equation}
and thus the approximation expected to be reliable at the typical frequency
(\ref{typicalfreq}) or  below
for $l_P^2/8\pi M^2 << 1$ i.e. for black holes of mass of a few Planck masses
or of higher mass.

\section{Conclusions}\label{conclusions}

In this paper we gave a detailed treatment of the late-time expansion which
occurs in the semiclassical approach to
the Hawking radiation. We find that the
variational problem defining the action related to the modes which are regular
at the horizon allows in general more than one solution, due to the presence
of caustics. We prove however that for radii below a critical value $r_c$
the variational problem has only one solution and we give a rigorous lower
bound on $r_c$. Thus for $r_1$ less that $r_c$ where the semiclassical
approximation is expected to be accurate there are non ambiguities in
computing the action and the time Fourier transform can be applied to extract
the Bogoliubov coefficients.  The Hamiltonian depends on the boundary
condition through a system of two highly non linear equations. 
We show that  for sufficiently late times such a system of equation is
rigorously equivalent to an other non linear equation which can be
solved through a convergent iterative procedure. We work out explicitely the
first two steps of such iteration which are sufficient to
compute the $\omega/M$ correction to the Hawking spectrum. The treatment shows
directly the relation between late times and high wave numbers of the modes 
regular at the horizon. The first two terms in the iterative process are
sufficient to give accurate results for the back-reaction effects for
frequencies at or below the typical frequency of the spectrum 
and black holes of a few Planck masses or higher mass.

\section*{Acknowledgments}

The author is grateful to Sergio Zerbini for useful discussions.

\section*{Appendix}  

We summarize here the essential formulas of the shell dynamics. For more
details see \cite{KWI,FLW,FM,menottiI}. One starts from the usual
Hilbert-Einstein action to which the shell action is added
\begin{equation}\label{specialF}
S=\frac{1}{16\pi G}\int R\sqrt{-g}~d^4x + S_{shell}.
\end{equation}
We shall in the following use $c=G=1$ which simply means that masses
acquire the dimension of length i.e. they are measured by the related
Schwarzschild radius divided by 2. As usual in gravity it is better to work
on a bounded region of space-time.
Employing the general spherically symmetric metric
(\ref{metric}) the action can be rewritten in Hamiltonian form as
\cite{FMP,KWI}
\begin{eqnarray}
S&=&\int_{t_i}^{t_f} dt 
\int_{r_i}^{r_e} dr (\pi_L \dot L +
\pi_R \dot R - N{\cal H}_t-N^r {\cal H}_r) +
\int_{t_i}^{t_f} dt \left.(-N^r \pi_L L+ \frac{NRR'}{L})\right|^{r_e}_{r_i}
\nonumber\\
&+&\int_{t_i}^{t_f} dt ~\hat p \dot{\hat r}
\end{eqnarray}
 where $\hat r$ denotes the radial coordinate of the shell. 
The constraints are given by
\begin{equation}
{\cal H}_r = \pi_R R'- \pi_L'L -\hat p~\delta(r-\hat r),
\end{equation}
\begin{equation}
{\cal H}_t = \frac{R R''}{L}+\frac{{R'}^2}{2
L}+\frac{L \pi_L^2}{2R^2}-\frac{R R' L'}{L^2}-
\frac{\pi_L\pi_R}{R}-\frac{L}{2}+ \sqrt{{\hat p}^2
L^{-2}+m^2}~\delta(r-\hat r).
\end{equation}
The Painlev\'e-Gullstrand gauge is defined by $L\equiv1$. 
There is still one gauge freedom in the choice of $R(r)$.
In virtue of the constraints $R'(r)$ has to be discontinuous
at $r=\hat r$. Here we will adopt the ``outer gauge'' \cite{FM}
defined by $R(r)=r$ for
$r\geq \hat r$ i.e. in the massless case
\begin{equation}
R(r) = r + \frac{\hat p}{\hat r}g(r-\hat r)
\end{equation}
with $g$ smooth function of support $[-l,0]$, $g(0)=0$ and $g'(0-)=1$.
Other gauges could well be used \cite{FM,menottiI}. 
The constraints can be solved and the action in the outer gauge 
takes the form
\begin{equation}
S = \int_{t_i}^{t_f} \left(p_{c}~ \dot{\hat r} -\dot M(t)\int_{r_i}^{\hat
r(t)}\frac{\partial F}{\partial M} dr - H N(r_e) + M N(r_i)\right)dt 
\end{equation}
where $F$ is the generating function
\begin{equation}
F=R W+ R R'({\cal L}-{\cal B})
\end{equation}
with
\begin{equation}
W = \sqrt{ R'^2 - 1 + \frac{2\cal M}{R}},~~~~ {\cal L}= \log(R'-W),~~~
{\cal B}=
\sqrt{\frac{2{\cal M}}{R}}+\log\bigg(1-\sqrt{\frac{2{\cal M}}{R}}\bigg).
\end{equation}
The general expression of the conjugate momentum $p_c$ is \cite{FM}
\begin{equation}
p_c = R(\Delta {\cal L} - \Delta{\cal B})
\end{equation}
where $\Delta$ represents the discontinuity of the related quantities across
the shell position $\hat r$.
Contrary to $\hat p$, $p_c$ is a gauge invariant quantity within the
Painlev\'e class of gauges \cite{menottiI}. 
Its expression for
the case of a massless shell is given by eq.(\ref{pc}).
Normalizing the lapse function $N$, which is constant for $r>\hat r$, as
$N(r_e)=1$ we have from the expression (\ref{pc}) of $p_c$  and action 
(\ref{reducedaction3}) the equation of motion
\begin{equation}
\frac{\partial H}{\partial p_c} = 1-\sqrt\frac{2H}{\hat r} =\dot{\hat r}
\end{equation}

\vfill


\end{document}